\documentstyle[12pt]{article}
\evensidemargin \oddsidemargin
\def\beq{\begin{equation}}
\def\eeq{\end{equation}}
\def\bea{\begin{eqnarray}}
\def\eea{\end{eqnarray}}
\def\ba{\begin{array}}
\def\ea{\end{array}}
\def\del{\partial}

\def\t{\widetilde}

\def\s{\sigma}

\def\al{\alpha}
\def\be{\beta}
\def\G{\Gamma}
\def\O{\Omega}
\def\e{\epsilon}
\begin{document}
\pagestyle{empty}
\begin{flushright}
hep-th/9907152\\
CPHT-S723.0799\\
\end{flushright}
\vspace*{.5cm}
\begin{center}
\vspace{.5cm}
{\Large\bf T-Duality, Space-time Spinors and R-R Fields in\\
\vspace{.5cm}
Curved Backgrounds}\\
\vspace*{1cm}
{\bf S. F. Hassan}{\footnote{\tt e-mail: fawad@cpht.polytechnique.fr}} \\
\vspace{.2cm}
{\it Centre de Physique Th\'{e}orique, Ecole Polytechnique, \\
91128 Palaiseau, France}\\ 
\vspace{1cm}
{\bf Abstract}
\begin{quote}
We obtain the T-duality transformations of space-time spinors (the
supersymmetry transformation parameters, gravitinos and dilatinos) of
type-II theories in curved backgrounds with an isometry. The
transformation of the spinor index is shown to be a consequence of the
twist that T-duality introduces between the left and right-moving
local Lorentz frames. The result is then used to derive the T-duality
action on Ramond-Ramond field strengths and potentials in a simple
way. We also discuss the massive IIA theory and, using duality, give a
short derivation of ``mass''-dependent terms in the Wess-Zumino
actions on the D-brane worldvolumes. 
\end{quote}
\end{center}
\vfill
\begin{flushleft}
{July 1999}\\
\vspace{.5cm}
\end{flushleft}
\newpage
\setcounter{footnote}{0}
\pagestyle{plain}
\section{Introduction}
The action of a T-duality transformation on the string worldsheet
fermions can be studied by demanding compatibility with the $N=1$
worldsheet supersymmetry. This determines the T-duality transformation
of the worldsheet spinors both in flat space \cite{DHS} as well as in
the presence of background fields with an isometry along which duality
is performed \cite{SFH1}. The effect on extended worldsheet
supersymmetry has been studied in \cite{IKR,SFH2}, and also in 
\cite{SFH1,BS} when the extended supersymmetry does not respect the
isometry. In this paper, we study the action of T-duality on
space-time fermions in Type-II superstring theories with background
fields and use the results to give a simple derivation of the R-R
T-duality rules.

In flat backgrounds, the action of T-duality on space-time spinors
follows in a rather straightforward way from its action on the
worldsheet currents $\del_\pm X^M$ and worldsheet fermions
$\psi_\pm^M$ \cite {DHS,Pol,Bach}: A T-duality with respect to $X^9$
sends $\del_+X^9 \rightarrow -\del_+X^9$ and $\psi_+^9 \rightarrow
-\psi_+^9$, keeping all other variables unchanged. Hence, on the
left-moving half of the worldsheet theory, it can be regarded as a
parity reflection along $X^9$, while the right-moving sector remains
invariant. The action of such a transformation on the left-moving
Ramond ground state is represented by $\O_0=\G_{11}\G^9$ as this
operator sends $\G^9$ of the left-moving sector to $\O_0^{-1}\G^9\O_0
=-\G^9$, consistent with the fact that $\G^9$ is the zero mode of
$\psi_+^9$ in the Ramond sector. The action of $\O_0$ can now be
absorbed in the space-time spinors. For example, for the parameters of
space-time supersymmetry $\e_\pm$ (where the subscripts ``$\pm$''
refer to the worldsheet sectors in which the supersymmetry acts), this
leads to $\e_+\rightarrow \O_0\e_+$, while $\e_-$ remains
unchanged. One can also obtain the action of T-duality on the
gravitinos $\Psi_{\pm M}$ from the invariance of their vertex
operators under T-duality. $\Psi_{+M}$ contains the left-moving R
ground state and hence transforms as $\Psi_{+M}\rightarrow
\O_0\Psi_{+M}$ while $\Psi_{-M}$ contains a left-moving NS field
$\psi_-^M$ and hence $\Psi_{-9}\rightarrow -\Psi_{-9}$,
$\Psi_{-i}\rightarrow\Psi_{-i}$.

In the general case of non-flat backgrounds with an isometry, say,
along $X^9$, T-duality no longer reduces to a parity transformation
acting on left-moving (or right-moving) worldsheet variables alone. In
fact, in general, it acts as a canonical transformation affecting both
left and right moving sectors of the worldsheet theory
\cite{GRV,AEL,SFH1}. Furthermore, in curved backgrounds, the 
relationship between the worldsheet fermions and space-time Dirac
algebra is not as straightforward as in flat space. Therefore, it does
not seem possible to obtain the T-duality action on space-time fermions,
or equivalently, on the Ramond sector, from worldsheet considerations
alone. 

In this paper, we study the action of T-duality on space-time spinors
in type-II string theories in the presence of NS-NS and R-R background
fields. The spinors we consider are the space-time supersymmetry
transformation parameters $\e_\pm$, the two gravitinos $\Psi_{M\pm}$
and the two dilatinos $\lambda_{\pm}$. These results are then used to
derive the T-duality rules for the R-R fields, including the massive
IIA case. Both backgrounds and spinors are assumed to be independent of
the coordinate $X^9$ along which T-duality is performed (with the
exception of type-IIB potentials dual to massive type-IIA).

The paper is organized as follows: In section 2, the transformation of
$\e_\pm$ is obtained by identifying a T-duality action on the local
Lorentz frame associated with the left-moving sector of the worldsheet
theory. We also describe a set of variables in terms of which the
curved-space duality resembles the flat-space case. In section 3, we
consider space-time supersymmetry transformations in type-II theories
in NS-NS backgrounds and determine the gravitino and dilatino
T-duality transformations. These are shown to be independent of R-R
backgrounds. In section 4, we use these transformations to derive the
T-duality rules for R-R field strengths and potentials, emphasizing
the locality of potentials in the massive type-IIA case. We then use
T-duality to give a simple derivation of the ``mass''-dependent terms
in the Wess-Zumino action for D-branes in massive IIA theory. Section
5 contains the conclusions. Many of the formulas used in this paper
are given in the appendix for convenience and to insure consistency of
conventions. 

\section{Representation of T-duality on Spinors in Curved Backgrounds}
In this section we describe how T-duality acts on the spinorial index
of space-time fermions in type-II theories with background fields (the
extension to other string theories is straightforward). This fully 
determines the transformation of the supersymmetry transformation
parameters $\e_\pm$.  

The action of T-duality on massless NS-NS sector fields $G_{MN}$, 
$B_{MN}$ and the dilaton $\phi$ is well known \cite{TB}. For later
reference, we write the result here in our T-duality conventions,
\bea
\t G_{9 9} &=& G^{-1}_{99}\,,
\nonumber\\
\t G_{9 i} &=&-G^{-1}_{99}B_{9 i}\,,
\nonumber\\
\t B_{9 i} &=&-G^{-1}_{99}G_{9 i}\,,
\nonumber \\
\t G_{i j} &=& G_{ij}-G^{-1}_{99}(G_{9 i}G_{9 j}-B_{9 i}B_{9 j})\,,
\nonumber \\
\t B_{i j} &=& B_{ij}-G^{-1}_{99}(G_{9 i}B_{9 j}-B_{9 i}G_{9 j})\,,
\nonumber\\
2\,\t\phi  &=& 2\,\phi - \ln G_{99}\,.
\label{dual}
\eea
Here, $M,N$ are space-time indices in 10 dimensions. The backgrounds
are assumed to be independent of the $X^9$ coordinate along which
T-duality is performed, but may depend on the remaining coordinates
which we label by $X^i$ with $i=0,1,\cdots,8$. Throughout this paper,
a tilde denotes a field in the T-dual theory. 

Let us decompose the 10 dimensional metric of type-II theories in
terms of the vielbeins, $G_{MN}=e_M^{\,\,a}\eta_{ab}e^{b}_{\,\,N}$, 
where $a,b$ are Lorentz frame indices. It is known that the T-dual
theory contains two possible vielbeins that we denote by $\t
e^M_{(-)a}$ and $\t e^M_{(+)a}$, both giving rise to the same T-dual 
metric $\t G^{MN}$ \cite{SFH1,SFH2,BEK}. Explicitly,  
\beq
\t e^M_{(-)a} = Q^M_{-N} e^N_{\,\,a}\,,\qquad
\t e^M_{(+)a} = Q^M_{+N} e^N_{\,\,a}\,.
\label{epm}
\eeq
The matrices $Q_{\pm}$ that implement T-duality on the vielbeins are
given (along with their inverses) by,   
\beq
Q_\pm =\left(
\ba{cc}\mp G_{99} & \mp(G\mp B)_{9 i} \\  & \\
             0    &  {\bf 1}_{9}
\ea\right)\,,
\qquad
Q_\pm^{-1} = \left(
\ba{cc}
\mp G^{-1}_{99} & -G^{-1}_{99}(G\mp B)_{9 i} \\ & \\
          0     &  {\bf 1}_{9}
\ea\right) \,,
\label{Qpm}
\eeq
where ${\bf 1}_{9}$ denotes the identity matrix in nine dimensions.
The two vielbeins in the dual theory are related by a local Lorentz
transformation $\Lambda^a_{\,\,b}$, 
\beq
\t e^M_{(+)b}= \t e^M_{(-)a} \Lambda^a_{\,\,b}\,,\qquad 
\Lambda=e^{-1}Q^{-1}_-Q_+e\,. 
\label{LTV}
\eeq 
Using the expressions for $Q_{\pm}$, it is easy to see that the matrix
$\Lambda$ is given by  
\beq
\Lambda^a_{\,\,b}= \delta^a_{\,\,b}-2G^{-1}_{99}e^a_{\,\,9}e_{9b}\,.
\label{Lambda}
\eeq
Note that $det\,\Lambda=-1$.
 
The appearance of two possible vielbeins in the dual theory is not an
inconsequential ambiguity and disregarding either of them will lead to
an inconsistent theory. In fact, it forces us to augment T-duality
with a local Lorentz transformation acting only on the Lorentz frame 
associated with the left-moving sector of the worldsheet theory.  To
see this, it is useful to regard the two vielbeins in
$G_{MN}=e_M^{\,\,a}\,\eta_{ab}\,e^{b}_{\,\,N}$ as the wavefunctions
associated with the left-moving and right-moving worldsheet operators,
respectively, that contribute to the graviton vertex operator. Though
these vielbeins may be assigned to different worldsheet sectors, they
are identical from the point of view of space-time geometry which does
not directly see the string worldsheet.  However, T-duality acts
differently on the two worldsheet sectors and one may expect it to
transform the corresponding vielbeins in different ways \footnote{That
T-duality could transform the vielbeins associated with the left- and
right-moving worldsheet sectors in different ways, is not evident from
the transformation of the metric.  This is because the T-duality
action on the metric is determined by the invariance of the
energy-momentum tensor and not that of the worldsheet
Lagrangian.}. That this is the origin of the difference between $\t
e_{(+)}$ and $\t e_{(-)}$ can be argued as follows: The left-moving
and right-moving worldsheet sectors are interchanged under the
worldsheet parity transformation $\sigma\rightarrow -\sigma$ which
also interchanges $Q_+$ and $Q_-$ \cite{SFH2} and, hence, the two
vielbeins in the dual theory. This suggests that $\t e^M_{(+)a}$ is
T-dual to the vielbein associated with the left-moving sector of the
original worldsheet theory, while $\t e^M_{(-)a}$ is T-dual to the
one associated with the right-moving sector. This identification also
gives a heuristic understanding of the T-duality action (\ref{epm}) on
the vielbeins: Note that in flat space, $e^a_M$'s appear as
wavefunctions for states created by Fourier modes of the worldsheet
fermions $\psi^M_{\pm}$. In curved backgrounds, T-duality transforms
these fermions to $\t\psi^M_\pm = Q^M_{\pm N}\psi^N_\pm$ \cite{SFH1},
which is consistent with the mapping of their associated wavefunctions
to $\t e_{(\pm)}$, as given by (\ref{epm}), depending on the worldsheet
sector they come from.

The necessity of retaining both $\t e_{(+)}$ and $\t e_{(-)}$ in the
dual theory is not evident if we are dealing with bosonic fields
alone. However, their presence is essential to insure the consistency
of the dual theory in the presence of space-time fermions, as will be
seen in the next section\footnote{That both vielbeins necessarily
appear in the dual theory also follows from the T-duality action on
complex structures associated with extended worldsheet supersymmetry,
in cases where the complex structures could be constructed in terms of
target-space Killing spinors (for example, in non-compact Calabi-Yau
in 4-dimensions \cite{KKL}) as discussed in \cite{SFH1}.}. This
implies that we have to keep track of how the vielbeins transform,
depending on the worldsheet sector they originate in. Then, to
reconcile the results with the standard formulation of gravity with
one set of vielbeins, we should use (\ref{LTV}) to re-express one of
the vielbeins, say $\t e_{(+)}$ in terms of the other one, {\it i.e.},
$\t e_{(-)}$. In other words, we have to augment the T-duality action
on the left-moving vielbein by a local Lorentz transformation, 
$e\rightarrow Q_+e\Lambda^{-1}$, so that it transforms to $\t e_-$,
rather than to $\t e_+$. This translates to the T-duality action on
the spinor index that the left-moving Ramond sector contributes to the
space-time fields. Formulating the dual theory in terms of $\t
e_{(-)}$ is natural since for self-dual backgrounds, $Q_{-}$ in
(\ref{Qpm}) reduces to the identity matrix and $\t e_{(-)}=e$ without
further field redefinitions (Though this is not the case with $\t
e_{(+)}$, choosing it will also lead to a physically equivalent
description). 

Consider the space-time supersymmetry transformation parameters
$\e_{\pm}$ and the Dirac matrices $\G^M=e^M_{\,\,a}\G^a$ in either IIA
or IIB theory. The Majorana-Weyl spinors $\e_\pm$ are taken to be
independent of $X^9$ and the subscripts ``$\pm$'' refer to their
worldsheet origin and not their space-time chirality which will depend
on the theory and will be specified later. After T-duality, we will
have two possible sets of $\G$-matrices,
\beq
\t\G^M_{(+)}=\t e^M_{(+)a}\G^a\,,\qquad \t\G^M_{(-)}=
\t e^M_{(-)a}\G^a\,.
\label{Gpm}
\eeq
Keeping track of their worldsheet origin, the spinors $\e_\pm$ in the
dual theory are associated with the Dirac algebras generated by
$\t\G^M_{(\pm)}$, respectively. The two sets of Dirac matrices are 
related by, 
\beq
\t\G^M_{(+)}=\O^{-1}\t\G^M_{(-)}\O\,,\qquad {\rm with}\,,\qquad
\O^{-1}\G^a\O = \Lambda^a_{\,\,b} \G^b\,.
\label{OL}       
\eeq
Clearly, $\O$ is the spinorial representation of the Lorentz
transformation (\ref{LTV}). The form of $\O$, including its
normalization, can be determined by the following argument:
Let us write the $\Lambda^a_{\,\,b}$ in (\ref{Lambda}) as     
\beq
\Lambda^a_{\,\,b}= \delta^a_{\,\,b}-2\omega^a_{\,\,b}\,,\qquad
{\rm with}\,, \qquad \omega^a_{\,\,b}=G^{-1}_{99}e^a_{\,\,9}e_{9b}\,.
\label{omega}
\eeq
One can easily verify that $\omega^a_{\,\,b}\omega^b_{\,\,c}=
\omega^a_{\,\,c}$,  so that $\omega^a_{\,\,b}$ is a projection operator
of rank 1. The operator $\omega=\omega^a_{\,\,b}(\del/\del X^a)dX^b$
projects the vector $\G=\G^a\del/\del X^a$ along the isometry generator 
$K$ which, normalized to unity, is given by $K=G_{99}^{-1/2}e^a_{\,\,9}
(\del/\del X^a)$. The projected component of $\G$ is then given by
$<K\,,\omega\cdot\G>$. The transformation $\Lambda^a_{\,\,b}$ in 
(\ref{omega}) changes the sign of this component, keeping other
components of $\G$ unchanged. Therefore, its spinor representation
$\O$ is obtained by multiplying the projected component with $\G_{11}$,
\beq
\O\,=\,\G_{11}<K\,,\omega\cdot\G>\,=\,\sqrt{G^{-1}_{99}}\G_{11}\G_9\,,
\label{LTS}
\eeq
as can be directly verified using (\ref{OL}). The sign of $\Omega$ is
not fixed by these considerations and its arbitrariness gives rise to
different T-duality conventions as we will discuss later. Note the
appearance of $\G_9=G_{9M}\G^M$ rather than $\G^9$ (as a naive
generalization from the flat-space case may suggest) in this
formula. This is related to the fact that, unlike in flat backgrounds
(or more generally, self-dual backgrounds defined by $Q_-=1$),
T-duality now mixes $\del_\pm X^9$ with other coordinates $\del_\pm
X^i$ when regarded as a canonical transformation in the worldsheet
theory. As will be apparent in section 4, The factor
$\sqrt{G^{-1}_{99}}$ in (\ref{LTS}) is essential for giving the
correct dilaton transformation, though here its existence was dictated
by different considerations.  

To write the dual theory with a single Dirac algebra basis, we express
$\t\G^M_{(+)}$ in terms of $\t\G^M_{(-)}$ using (\ref{OL}), and absorb
$\O$ in a redefinition of the spinor $\e_+$, with $\e_-$ remaining
unchanged. This gives the T-duality transformation rules for the
space-time supersymmetry parameters $\e_{\pm}$, which are the simplest
spinorial objects in the theory, as   
\beq
\ba{l}
\t\e_-=\,\e_-\,, \cr
\t\e_+=\,a_{(o-f)}\,\O\,\e_+\,,\qquad{\rm where}\,,
\qquad a_{(o-f)}=\pm 1\,.
\ea
\label{etd}
\eeq 
Note that $\t\e_+$ and $\e_+$ have opposite space-time chiralities,
which is the basis of IIA-IIB interchange under T-duality. The factor
$a_{(o-f)}$ (with ``{\it o}'' standing for {\it original} and ``{\it
f}'' for {\it final}) reflects the arbitrariness in the sign of $\O$.
It is used to denote $a_{(A-B)}$ when T-duality converts an original
IIA theory to a final IIB theory, and $a_{(B-A)}$ {\it vice versa}.
The arbitrariness in sign allows for two distinct conventions:
Consider two successive T-duality transformations along $X^9$. Since
$\t\O=\O$, as can be verified using (\ref{dual}), we have $\t\O
\O=-1$. If we choose the convention $a_{(A-B)}=a_{(B-A)}$, then
$\t{\t\e}_+ = -\e_+$. In fact, with this convention, all left-moving
Ramond states behave in this way and T-duality squares to $(-1)^{F_L}$
on the spectrum (where $F_L$ is the left-moving space-time fermion
number). However, since IIA and IIB are different theories, we can
also choose the alternative convention,
\beq
a_{(A-B)}=-a_{(B-A)}\,,
\label{aof}
\eeq
in which case, the T-duality operation that takes IIA to IIB is the
inverse of the one that takes IIB to IIA, and the transformation
squares to $+1$ on the spectrum. In the following, we use the latter
convention whenever a convention is explicitly specified. The
correctness of equations (\ref{etd}) will be checked in the next
section when we examine the supersymmetry variations of gravitinos and
dilatinos to extract their T-duality transformations.

Unlike the flat-space case, in non self-dual backgrounds the
canonical transformation that implements T-duality acts on both
worldsheet sectors. Explicitly \cite{SFH1}, 
\bea
&&\t\psi^M_\pm = Q^M_{\pm N} \psi^N_\pm\,,\nonumber\\
&&\del_\pm\t X^M = Q^M_{\pm N} \del_\pm X^N + \psi^i_\pm\del_i
Q^M_{\pm N} \psi^N_\pm\,.
\label{ct}
\eea
These equations are non-trivial only for $\del_\pm\t X^9$ and
$\t\psi^9_\pm$, reducing to $\t X^i=X^i$ and $\t\psi_\pm^i=\psi_\pm^i$
for the rest. However, the invariance of $\e_-$ may tempt one to
search for variables in terms of which T-duality in curved space has
the same form as that in flat-space, affecting only the left moving
sector. To find such variables, note that the matrices $Q_{\pm}$,
which have a very simple upper triangular form, may be decomposed as
$$
Q_+={\t A_+}^{-1}\left(\ba{cc}-1&0\\0&{\bf 1}_9\ea\right) A_+\,,
\qquad\qquad
Q_- = {\t A_-}^{-1} A_-\,.
$$
Here $\t A_\pm$ are the same matrices as $A_\pm$, but in the dual
theory. These equations admit many solutions, all with
$A^i_{\pm N}=\delta^i_N$ while $A^9_{\pm M}$ are not uniquely
determined (for example, $A^9_{-M}=A^9_{+M}=G_{9M}/\sqrt{G_{99}}$). If
we define new worldsheet fermionic and bosonic variables,  
\bea
\Sigma_\pm^M &=& A^M_{\pm N} \psi_\pm^N \,,\nonumber\\ 
J_\pm^M &=& A^M_{\pm N}\del_\pm X^N+\psi^j_\pm\,\del_j A^M_{\pm N}
\,\psi_\pm^N\,,
\nonumber
\eea
then the canonical transformations (\ref{ct}) implementing T-duality
take the flat-background form,  
$$
\t J^9_+=-J^9_+ \,,\qquad\qquad \t\Sigma^9_+=-\Sigma^9_+\,,
$$
with $J^9_-$ and $\Sigma^9_-$ unchanged ($J^i_\pm=\del_\pm X^i$ and 
$\Sigma^i_\pm=\psi^i_\pm$ are trivially invariant). However, the
Lagrangian in terms of the new variables does not look any simpler
which shows the basic difference between the self-dual ($Q_-=1$),
and the more general non self-dual cases, even though the
transformations can be written is a similar form.   

\section{Action of T-duality on Gravitinos and Dilatinos}
In this section we will derive the transformations of the type-II
superstring gravitinos $\Psi_{\pm M}$ (not to be confused with the
worldsheet spinors $\psi_\pm^M$) and dilatinos $\lambda_{\pm}$ under
T-duality, by demanding compatibility between T-duality and space-time
supersymmetry. Again, the ``$\pm$'' subscripts refer to the worldsheet
sector in which the spinor index of the fermion, {\it i.e.} its Ramond
component, originates and not to its space-time chirality. All spinors
are assumed to be independent of the coordinate $X^9$ along which
T-duality is performed. The T-duality action on these spinors is
independent of the R-R fields, which we set to zero in this
section for convenience. The case of non-zero R-R fields will be
considered in the next section.

Let us first consider the supersymmetry variations of the gravitinos
$\Psi_{\pm M}$. With $\e_{\pm}$ as the supersymmetry transformation
parameters and in the absence of R-R fields, these are given by 
\bea
\delta_\pm \Psi_{\pm M}&=&\left(\del_M + \frac{1}{4}\,W^\pm_{Mab}\,
\G^{ab}\right)\e_\pm\,\,+\,\cdots \,,
\label{dG-NS}\\
\delta_\pm \Psi_{\mp M}&=& 0\,\,+\cdots\,.
\label{dG-R}
\eea
Here, ``$\cdots$'' indicates the presence of 3-spinor terms that we
do not write down explicitly, but which will be automatically accounted
for in our final result. $W^\pm_{Mab}$ are the torsionful
spin-connections given by 
\beq
W^\pm_{Mab}=w_{Mab}\mp\frac{1}{2}H_{Mab}\,.
\eeq 
The above transformations hold in both IIA and IIB theories, depending
on the chirality of the spinors.  In our conventions, in IIB, $\e_\pm$
and hence $\Psi_{\pm M}$ have positive chirality while in IIA, $\e_-$,
$\Psi_{-M}$ have positive chirality and $\e_+$, $\Psi_{+M}$ have
negative chirality. The supersymmetry transformation generated by
$\e_+$ ($\e_-$) acts on the left-moving (right-moving) worldsheet
sector by interchanging R and NS boundary conditions. Therefore, the
supersymmetry variations $\delta_\pm \Psi_{\pm M}$ convert R-NS
states into NS-NS states and do not get modified if R-R fields are
switched on. Therefore, we expect that the gravitino T-duality rules
obtained from equation (\ref{dG-NS}) are independent of R-R
fields. The same argument applies to dilatino T-duality rules.  

Let us now consider the gravitino supersymmetry variations in the
T-dual theory. First, note that the dual theory contains two sets of
torsionful spin-connections, corresponding to the two vielbeins $\t
e^a_{(-)M}$ and $\t e^a_{(+)M}$ given by (\ref{epm}). We denote these
by $\t W^\pm_{(-)Mab}$ and $\t W^\pm_{(+)Mab}$, respectively. One can 
verify that
\bea
\t W^-_{(-)Mab} &=& W^-_{Nab} (Q^{-1}_+)^N_{\,\,M}\,, 
\label{W--}\\
\t W^+_{(+)Mab} &=& W^+_{Nab} (Q^{-1}_-)^N_{\,\,M}\,.
\label{W++}
\eea
Since we have chosen to express the T-dual theory in terms of $\t
e_{(-)}$, the supersymmetry variations $\delta_\pm\t\Psi_{\pm M}$ 
in the T-dual theory should be expressed in terms of $\t 
W^\pm_{(-)Mab}$ alone,
\beq
\delta_\pm \t\Psi_{\pm M}=\left(\del_M + \frac{1}{4}\t W^\pm_{(-)Mab}
\G^{ab}\right)\t\e_\pm\,\,+\,\cdots \,.
\label{dTDG-NS}
\eeq
To determine $\t\Psi_{+ M}$ in terms of $\Psi_{+ M}$, note that using
the relation (\ref{LTV}) between $\t e_{(+)}$ and $\t e_{(-)}$, we can
write $\t W^+_{(+)Mab}$ in terms of $\t W^+_{(-)Mab}$ as,   
\bea
\t W^{+\,\,\,\,\,\,\,\,\,\,a}_{(+)M\,\,b}\G_a^{\,\,\,b}
&=&\t W^{+\,\,\,\,\,\,\,\,\,\,c}_{(-)M\,\,d}\,
(\Lambda^{-1})^a_{\,\,c}\,\Lambda^d_{\,\,b}\,\G_a^{\,\,\,b}
+ (\Lambda^{-1})^a_{\,\,c}\del_M \Lambda^c_{\,\,b}\,\G_a^{\,\,\,b}
\nonumber\\
&=&\t W^{+\,\,\,\,\,\,\,\,\,\,a}_{(-)M\,\,b}\O^{-1} \G_a^{\,\,\,b}\O
+4\O^{-1}\del_M\O\,.
\label{W+W-}
\eea
Now, using equations (\ref{etd}) and (\ref{W--}-\ref{W+W-}), along
with the fact that $Q_{\pm j}^i=\delta^i_{\,\,j}$ (\ref{Qpm}) and
$\del_9\e_- =0$, it is easy to see that the variation (\ref{dG-NS})
implies the one in the dual theory (\ref{dTDG-NS}) provided, 
\bea
\delta_-\t\Psi_{-M}&=&\delta_-\Psi_{-N}\,(Q^{-1}_+)^N_{\,\,M}\,+\,
\cdots\,,
\label{TDdG-}\\
\delta_+\t\Psi_{+M}&=&a_{(o-f)}\,\O\,\delta_+\Psi_{+N}\,
(Q^{-1}_-)^N_{\,\,M}\,+\,\cdots\,.
\label{TDdG+}
\eea
Again, ``$\cdots$'' denotes 3-spinor terms.

Let us now consider the supersymmetry variations of the dilatinos
$\lambda_\pm$ in the absence of R-R fields, 
\bea
\delta_\pm \lambda_\pm&=& \frac{1}{2}\left(\G^M\del_M \phi \mp 
\frac{1}{12}\G^{MNK} H_{MNK}\right)\e_\pm\,\,+\cdots\,,
\label{dD-NS}\\ 
\delta_\pm \lambda_\mp &=&0\,\,+\cdots\,.
\label{dD-R}
\eea
These are again valid in both IIA and IIB theories. In IIB, both
dilatinos have negative chirality, while in IIA, $\lambda_-$ has
negative chirality and $\lambda_+$ has positive chirality. Switching
on R-R fields does not affect equation (\ref{dD-NS}). In the T-dual
theory, written in terms of the vielbein $e^a_{(-)M}$, the variations
$\delta_\pm\t\lambda_\pm$ are given by  
\beq
\delta_\pm \t\lambda_\pm=\frac{1}{2}\left(\t\G^M_{(-)}\del_M\t\phi 
\mp\frac{1}{12}\t\G^{MNK}_{(-)}\t H_{MNK}\right)\t\e_\pm\,\,+\cdots\,.
\label{dTDD-NS} 
\eeq
Using $\t\phi=\phi-\frac{1}{2}\ln G_{99}$ and 
\beq
\t\G^{MNK}_{(\mp)}\t H_{MNK}=\G^{MNK}H_{MNK}\mp 6\,G^{-1}_{99}\,\G_9
\,(W^\mp_{9ab}\,\G^{ab}) \pm 6\, G^{-1}_{99}\,\G^i\,\del_i\, G_{99}\,,
\label{TDGH}
\eeq
along with equation (\ref{etd}), one can see that the supersymmetry
variations (\ref{dD-NS}) and (\ref{dTDD-NS}) are compatible provided 
\bea 
\delta_-\t\lambda_- &=& \delta_-\lambda_- -G^{-1}_{99}\,\G_9\,\delta_-
\Psi_{-9}\,+\cdots\,,
\label{TDdD-}\\
\delta_+\t\lambda_+ &=& a_{(o-f)}\,\O\,\left(\delta_+\lambda_+
-G^{-1}_{99}\,\G_9\,\delta_+\Psi_{+9}\right)\,+\cdots\,.
\label{TDdD+}
\eea

Equations (\ref{TDdG-}),(\ref{TDdG+}) and (\ref{TDdD-}),(\ref{TDdD+})
give the T-duality transformations of the supersymmetry variations
$\delta\Psi_{\pm M}$ and $\delta \lambda_\pm$ to linear order in
spinors and receive corrections cubic in the spinors whose presence is
indicated by ``$\cdots$''. From these we can read off the T-duality
transformations of the gravitinos and dilatinos, in principle, only to
linear order in the spinors. However, as we will show, the linear
order result is exact and in fact, it dictates the form of the
3-spinor corrections to the T-duality maps for the supersymmetry
variations above. Thus, for the gravitinos $\Psi_{\pm M}$, we have the
T-duality transformations
\beq
\ba{rcl}
\t\Psi_{-M} & = &\Psi_{-N}\,(Q^{-1}_+)^N_{\,\,M}\,,\\ 
\t\Psi_{+M} & = & a_{(o-f)}\,\O\,\Psi_{+N}\,(Q^{-1}_-)^N_{\,\,M}\,, 
\ea
\label{TDG}
\eeq
and for the dilatinos $\lambda_\pm$ we have the transformations
\beq
\ba{rcl}
\t\lambda_- &=& \lambda_- -G^{-1}_{99}\,\G_9\,\Psi_{-9}\,,\\ 
\t\lambda_+ &=& a_{(o-f)}\,\O\left(\lambda_+-G^{-1}_{99}\,
\G_9\,\Psi_{+9}\right)\,.
\ea
\label{TDD}
\eeq
Here, $\O$ is given by (\ref{LTS}) and, as described below equation
(\ref{etd}), $a_{(o-f)}=\pm 1$ stands for $a_{(A-B)}$ if T-duality
takes us from IIA to IIB, and for $a_{(B-A)}$ if it acts the other way
round. Setting $a_{(A-B)}=-a_{(B-A)}$ insures that T-duality squares
to $+1$ on the spectrum. 

That equations (\ref{TDG}) and (\ref{TDD}) do not receive corrections
can be seen as follows: To linear order in spinors, these equations
are uniquely determined by (\ref{TDdG-}), (\ref{TDdG+}) and
(\ref{TDdD-}), (\ref{TDdD+}) thus only leaving the possibility of
adding corrections cubic in the spinors. The presence of such terms,
however, can be ruled out on general grounds as they would give rise
to derivative interactions for the spinors in the dual supergravity
action. To rule out, in a more concrete way, the existence of both
3-spinor corrections, as well as corrections proportional to R-R
fields, we consider the supersymmetry variations of the NS-NS fields
$G_{MN}$, $B_{MN}$ and $\phi$ given by equations (\ref{ap-dNS}) in the
appendix. These variations contain no R-R fields and are only bilinear
in spinors. Using (\ref{dual}) along with (\ref{TDG}) and (\ref{TDD})
one can easily verify that that these variations are consistent with
T-duality. On the other hand, if (\ref{TDG}) and (\ref{TDD}) contained
either 3-spinor terms or R-R dependent terms, this would not be the
case. This establishes that the spinor T-duality rules given above are
exact. Note that the NS-NS supersymmetry variations (\ref{ap-dNS}) are
insensitive to the multiplicative factor $\O$. Therefore, while they
can be used to rule out additional additive contributions to
(\ref{TDG}) and (\ref{TDD}), they cannot be used to infer the
existence of $\O$ in these transformations.

For supersymmetric backgrounds, when the fermionic backgrounds
$\Psi_\pm$ and $\lambda_\pm$ along with their supersymmetry variations
are set to zero, equations (\ref{dG-NS}) and (\ref{dD-NS}) reduce to
the string theoretic Killing spinor equations for $\e_\pm$. Equations
(\ref{TDG}) and (\ref{TDD}) are then trivial for the background
spinors, but can be used to obtain the T-duality transformation of the
fermionic excitations around supersymmetric backgrounds. In some
cases, when the Killing spinor itself does not transform (as is
the case with $\e_-$), the compatibility of the Killing spinor
equation with T-duality was investigated in \cite{BEK,BKO}. 

\section{R-R T-duality Revisited}

As shown above, the T-duality rules for space-time fermions do not
depend on the R-R fields. In this section we use these rules, along
with the requirement of compatibility of T-duality with space-time
supersymmetry, to determine the T-duality rules for R-R fields and
discuss some related issues. Most of the results in this section are
not new but are re-derived here in a unified and more convenient way.
The IIA/IIB T-duality rules for R-R fields were derived in
\cite{BHO,BRGPT} by studying the supergravity action and equations of
motion (also see \cite{EJL}) and in \cite{BR,GHT,Simon} by dimensional
reduction of the Wess-Zumino term in the D-brane worldvolume action,
both considering the bosonic sector alone. Our derivation of these
rules here emphasizes the compatibility of the T-duality conventions
used for the R-R fields with those used for the spinors. The T-duality
rules relating IIB to the massive IIA theory were obtained in
\cite{BRGPT,BR,GHT}. Here we re-derive these rules for generic
configurations, emphasizing how potential non-localities in the
T-duality rules for R-R potentials are avoided. We also present a
simple derivation of the ``mass''-dependent terms in the Wess-Zumino
action for ``massive'' IIA branes using T-duality.

In the presence of R-R backgrounds, the supersymmetry variations
$\delta_+\Psi_{+M}$ and $\delta_-\Psi_{-M}$ are still given by
(\ref{dG-NS}), while $\delta_-\Psi_{+M}$ and $\delta_+\Psi_{-M}$
are no longer zero and receive contributions from R-R fields. The
same is true for the dilatino variations $\delta_\pm\lambda_\mp$. The
T-duality rules for the R-R fields can be obtained by considering any
one of these variations, say $\delta_-\Psi_{+M}$. In type IIA theory,
this variation is given by \cite{LR} (see appendix C for details),
\bea
\delta_-\Psi_{+M}=\frac{1}{8}e^\phi\Big[F^{(0)}+\frac{1}{2!}\G^{M_1M_2}
F^{(2)}_{M_1M_2}+\frac{1}{4!}\G^{M_1M_2M_3M_4}F^{(4)}_{M_1M_2M_3M_4}
\Big]\G_M\e_- +\cdots\,,
\label{dG-RRA}
\eea
where ``$\cdots$'' denote 3-spinor terms as usual. $F^{(0)}=m$ is the
mass parameter of massive type-IIA theory and the field strengths
$F^{(n)}$ for the massive theory are given by (\ref{apD-Fm}) in the
appendix. The usual massless IIA equations are obtained by setting 
$m=0$. In type-IIB theory the corresponding variation is given by 
\cite{JHS} (see appendix B for details), 
\bea
\delta_-\Psi_{+M}=-\frac{1}{8}e^\phi\Big[\G^{M_1}F^{(1)}_{M_1}
\!\!\!&+&\!\!\!\frac{1}{3!}\G^{M_1M_2M_3}F^{(3)}_{M_1M_2M_3}
\nonumber\\&+&\!\!\!
\frac{1}{2(5!)}\G^{M_1M_2M_3M_4M_5}F^{(5)}_{M_1M_2M_3M_4M_5}\Big]
\G_M\e_- +\cdots\,.
\label{dG-RRB}
\eea
It is convenient to write these two equations in the generic form
\beq
\delta_-\Psi_{+M}=\frac{1}{2(8)}e^\phi \left[\sum_{n}
\frac{(-1)^n}{n!} \G^{M_1\cdots M_n} F^{(n)}_{M_1\cdots M_n}\right]
\G_M \e_- +\cdots\,.
\label{dG-RRAB}
\eeq
In exactly the same way as for the R-R vertex operator in flat
space (see, for example, \cite{Pol, Bach}), the actual content of the
above equation is determined by the chirality of the space-time
spinors: In type-IIB theory, both $\e_-$ and $\Psi_{+M}$ have positive
chirality and therefore the right hand side contains only terms with even
number of $\G$-matrices (corresponding to $n=1,3,5,7,9$), whereas in
IIA, $\e_-$ and $\Psi_{+M}$ have positive and negative chiralities
respectively and hence only terms with even $n$ ($n=0,2,4,6,8,10$)
enter the summation. Furthermore, using the $\G$-matrix identity
(\ref{apA-G2}), the positive chirality of $\e_-$ implies that 
$F^{(n)}=-(-1)^{n(n-1)/2}\, {}^*\!F^{(10-n)}$. This allows us to write
the summation in terms of $F^{(n)}$ with $n\le 5$ alone, recovering  
(\ref{dG-RRA}) and (\ref{dG-RRB}). 

Let us now consider the above equation in the T-dual theory expressed
in terms of the vielbein $\t e^a_{(-)M}$, 
\beq
\delta_-\t\Psi_{+M}=\frac{1}{2(8)}e^{\t\phi} \left[\sum_{n}
\frac{(-1)^n}{n!}\t\G_{(-)}^{M_1\cdots M_n}\t F^{(n)}_{M_1\cdots M_n}
\right]\t\G_{(-)M} \t\e_-+\cdots\,.
\label{dG-RRABTD}
\eeq 
Using equations (\ref{epm}), (\ref{Gpm}), (\ref{etd}) and (\ref{TDG}),
one can readily obtain the T-duality transformation for the R-R field
strengths as 
\beq
\t F^{(n)}_{M_1\cdots M_n}= (-1)^n a_{(o-f)} \left(F^{(n+1)}_{9N_1\cdots N_n}
+ n G_{9[N_1}F^{(n-1)}_{N_2\cdots N_n]}\right)(Q^{-1}_-)^{N_1}_{\,\,
M_1} \cdots (Q^{-1}_-)^{N_n}_{\,\,M_n}\,. 
\label{FQtd}
\eeq
where, $a_{(o-f)}$ denotes a convention dependent sign as explained
below equation (\ref{etd}). Let us now choose the convention
(\ref{aof}) so that T-duality squares to 1 on R-R fields. Then, using
the form of $Q^{-1}_-$ given in (\ref{Qpm}), the above equation
reduces to the component form, 
\bea
&&\t F^{(n)}_{9i_2\cdots i_n} = -a_{(A-B)}\left[ F^{(n-1)}_{i_2\cdots i_n}
-(n-1) G^{-1}_{99} G_{9[i_2} F^{(n-1)}_{9i_3\cdots i_n]}\right]\,,
\label{F1td}\\
&&\t F^{(n)}_{i_1i_2\cdots i_n} = -a_{(A-B)} F^{(n+1)}_{9i_1\cdots i_n}
-n B_{9[i_1} \t F^{(n)}_{9i_2\cdots i_n]}\,.
\label{F2td}
\eea
$a_{(A-B)}$ is still arbitrary and could be chosen as either $+1$ or
$-1$. The antisymmetrization denoted by the square bracket affects the
indices $i_n$ and not the index $9$. Since the spinors were assumed to
be independent of $X^9$, equation  (\ref{dG-RRAB}) implies that
$F^{(n)}$ should also be independent of this coordinate. 

The above T-duality rules for $F^{(n)}$ are valid for both massless
and massive type-IIA theories and can be iteratively integrated
to yield the corresponding transformations for the R-R potentials
$C^{(n)}$. Let us first consider duality between IIB and massless
IIA. In this case, $F^{(0)}=m=0$ and the field strengths are given by 
(\ref{apC-F}). $C^{(n)}$ can be chosen to be $X^9$-independent and
under T-duality transform as \cite{BHO,EJL,BRGPT,GHT,Simon}
\bea
&&\t C^{(n)}_{9i_2\cdots i_n}=a_{(A-B)}\left[ C^{(n-1)}_{i_2\cdots i_n}
-(n-1) G^{-1}_{99} G_{9[i_2} C^{(n-1)}_{9i_3\cdots i_n]}\right]\,,
\label{C1td}\\
&&\t C^{(n)}_{i_1i_2\cdots i_n}=a_{(A-B)} C^{(n+1)}_{9i_1\cdots i_n}
-n B_{9[i_1} \t C^{(n)}_{9i_2\cdots i_n]}\,.
\label{C2td}
\eea

Let us now consider the massive-IIA case. For $n=0$, equation
(\ref{F2td}) reduces to $\t F^{(0)}=-a_{(A-B)}F^{(1)}= -a_{(A-B)}
\del_9 C^{(0)}$. As noticed in \cite{BRGPT,BR,GHT}, this implies that
type-IIB theory dualizes to the massive IIA theory with $\t
F^{(0)}=m$, provided the IIB $0$-form has an $X^9$ dependence given by
$C^{(0)}= -a_{(A-B)}m X^9 + \hat C^{(0)}$, where the last term is
independent of $X^9$. Naively, one may expect that this
$X^9$-dependence could lead to a similar dependence for the IIA
potentials, which should not be the case: Consider an $X^9$-dependent
function $C(X^9)$, say, in the IIB theory leading to an
$X^9$-dependent T-dual $\t C(X^9)$ in IIA. Since the natural variable
in the T-dual theory is $\t X^9$, which is related to $X^9$ through
the canonical transformation (\ref{ct}), $\t C$ has to be expressed in
terms of $\t X^9$. However, the relationship between $X^9$ and $\t
X^9$ is non-local, involving an integration over the string
worldsheet, and hence $\t C$ is a non-local function of $\t X^9$. This
problem can be avoided if we arrange things such that the
$X^9$-dependent $C$ dualizes to an $X^9$-independent $\t C$, or {\it
vice versa}. Let us define
\bea
&&\widehat C^{(0)}=C^{(0)}+a_{(A-B)}m X^9 \,,
\nonumber\\
&&\widehat C^{(2)}_{M_1M_2}=C^{(2)}_{M_1M_2}+a_{(A-B)}mX^9 B_{M_1M_2}\,, 
\nonumber\\
&&\widehat C^{(4)}_{M_1\cdots M_4}= C^{(4)}_{M_1\cdots M_4} + 3a_{(A-B)} 
m X^9 B_{[M_1M_2} B_{M_3M_4]}\,, \nonumber
\eea
or more generally, using the notation of \cite{GHT}, with 
$C=\sum_{n=0}^9 C^{(n)}$, 
\beq
\widehat C = C+ a_{(A-B)}\,m\,X^9 e^B\,.
\label{CC}
\eeq
We give the $C^{(2p)}$ in type-IIB a dependence on $X^9$ in such a way
that $\widehat C^{(2p)}$ are $X^9$ independent, while in type-IIA,
$\widehat C^{(2p+1)}=C^{(2p+1)}$ and are $X^9$ indepentent. Then,
using the T-duality rules for the field strengths
(\ref{F1td},\ref{F2td}), along with equations (\ref{apC-F}) for
type-IIB and (\ref{apD-Fm}) for the massive type-IIA, one can obtain
the T-duality rules for the potentials. These are still given by
(\ref{C1td}) and (\ref{C2td}) but now with all $C^{(2p)}$ replaced by
$\widehat C^{(2p)}$. The $X^9$ independence of $\widehat C^{(2p)}$
guarantees the $X^9$ independence of the IIA potentials $C^{(2p+1)}$,
preventing the appearance of non-localities. Note that while the
massive T-duality rules, written in terms of $\widehat C^{(2p)}$, have
the same form as the usual massless IIA/IIB rules, the two differ by
$m$-dependent terms when written in terms of the actual R-R potentials
$C^{(2p)}$.

The special $X^9$ dependence of $C^{(2p)}$ can be easily understood 
when massive-IIA/IIB duality is regarded as a Scherk-Schwarz
compactification to 9 dimensions \cite{BRGPT}: The $U(1)\subset SL(2,R)$
transformation in IIB theory that gives the right $X^9$ dependence to
$C^{(0)}$, by shifting it to $C^{(0)}-a_{(A-B)}m X^9$ (corresponding to
$p=s=1$, $r=0$ and $q=-a_{(A-B)}m X^9$ in (\ref{apC-SL2R})), also
produces the correct $X^9$ dependences in $C^{(2)}$ and $C^{(4)}$. 

For $m\ne 0$, the Wess-Zumino terms in the IIA D-brane worldvolume
actions contain $m$-dependent terms the forms of which were studied
in \cite{BR,GHT}. We will now derive these terms in a very
straightforward way using T-duality: Let us start with the WZ terms in
the D-brane worldvolume actions in type-IIB theory and express the
potentials $C^{(2p)}$ in terms of $\widehat C^{(2p)}$ as defined in
(\ref{CC}),  
\beq
I^{(IIB)}_{WZ}=\int_{\omega_{2p}} C e^{F-B}= \int_{\omega_{2p}} 
{\widehat C}e^{F-B} - a_{(A-B)}m \int_{\omega_{2p}} X^9 e^F\,.
\label{IWZ}
\eeq 
When $C^{(2p)}$ are chosen such that $\widehat C^{(2p)}$ are $X^9$
independent, the dual theory is massive type-IIA. Therefore, on
dimensional reduction, $I^{(IIB)}_{WZ}$ should reduce to the
corresponding action for massive IIA theory, including the
$m$-dependent terms. As mentioned earlier, the massive T-duality rules
relating $\widehat C^{(2p)}$ and $C^{(2p+1)}$ have exactly the same
form as the massless T-duality rules relating $C^{(2p)}$ and
$C^{(2p+1)}$. Therefore, the analysis for the massless case, for
example, as presented in \cite{Simon} or \cite{GHT}, implies that the
first term on the right hand side of (\ref{IWZ}) dualizes to the
standard WZ term in type-IIA which is common between the massive and
massless theories. The $m$-dependent terms are contained in the second
integral on the right hand side of (\ref{IWZ}). Let us identify $X^9$
with a worldvolume direction, say $\sigma$, along which the theory is
reduced. Taking $F$ to be Abelian ($F={\rm d}V$), we write
$e^F=\sum_p\frac{1}{p!} {\rm d}(V\wedge F^{p-1})$ so that,
\beq
\int_{\omega_{2p}} X^9 e^F=\sum_p\frac{1}{p!}\int_{\omega_{2p}} X^9 
\left[\del_\s (V\wedge F^{p-1})\wedge{\rm d}\s +\del_\al (V\wedge
F^{p-1})\wedge{\rm d}x^\al\right]\,, 
\eeq
where, $x^\al$ are the worldvolume directions transverse to
$\s$. Since $X^9$ does not depend on $x^\al$, the second term in the
integrand leads to a surface term and can be dropped. The
$(2p-1)$-form $V\wedge F^{p-1}$ in the first term now only has
non-zero components along $x^\al$, and not along $\s$. Finally,
remembering that $X^9=\s$ and dropping a surface term, the integration
over $\s$ leads to
\beq 
I^{(IIA)}_{WZ}=\int_{\omega_{2p-1}} C e^{F-B} + 
a_{(A-B)} m \sum_p\frac{1}{p!}\int_{\omega_{2p-1}}V\wedge
({\rm d}V)^{p-1}\,, 
\eeq
which reproduces the m-dependent terms of \cite{BR,GHT} (we have
ignored the D-brane tension that can be easily inserted into the
equations). 
 
\section{Conclusions}
We have shown that, besides acting on the space-time indices of
fields, T-duality also has an action on the local Lorentz frame
associated with the left-moving sector of the worldsheet theory by
twisting it with respect to the one associated with the right-moving
sector. This twist translates to the T-duality action on the spinor
index originating in the left-moving Ramond sector, and fixes the
T-duality action on the space-time supersymmetry parameters. The
gravitinos and dilatinos also contain an NS sector contribution to
their T-duality transformations which is obtained by demanding
consistency between T-duality and space-time supersymmetry. It is also
shown that the T-duality action on the spinors is independent of the
R-R backgrounds.  The result is then used to re-derive the R-R
T-duality rules. We discuss the case of the massive IIA theory in more
detail, showing that there exist variables in terms of which the
massive T-duality rules for the R-R potentials have the same form as
the massless ones, manifestly avoiding non-local relations between
potentials. Using this, we give a simple derivation of the
``mass''-dependent terms in the WZ actions for the associated D-branes
based on T-duality. In most part, we have explicitly retained the
convention dependence of the T-duality action on the Ramond sector. In
one convention, T-duality squares to $1$, while in the other, it
squares to $(-1)^{F_L}$ on the spectrum, where $F_L$ is the left-moving
space-time fermion number. 

There are certain similarities between T-duality in flat and curved
backgrounds. At the worldsheet level, as we have shown, there exist
variables in terms of which the canonical transformation that
implements T-duality in curved space, has the flat-space form. One can
also check that the T-duality rules for gravitinos and R-R fields in
curved backgrounds easily follow from their flat-space vertex
operators, provided we interpret these operators as curved space
objects (which, of course, is not really the case). For example,
consider the gravitino emission operator $\sim{\bar S}_{+s}\Psi_{+M}^s 
\psi^M_-$ in flat space. To interpret this as a curved-space
expression, we define the spin-filed $S_{+s}$ as an operator that
generates space-time supersymmetry transformations of $\Psi_{+M}^s$
with parameter $\e_+^s$, but now in curved-space. $S_+$ and $\e_+$
will have opposite space-time chiralities and $\bar S_+\e_+$ is
invariant under T-duality. Then using the curved-space T-duality rules
for $\e_+$ (\ref{etd}) and $\psi^M_-$ (\ref{ct}) in the flat-space
vertex operator, we recover the T-duality action (\ref{TDG}) on
$\Psi_{+M}$. Similarly, the R-R T-duality rules can be obtained from
the corresponding flat-space vertex operator, $e^\phi\bar S_{+s}
F^{ss'} S_{-s'}$, where $F^{ss'}$ is the R-R bi-spinor.

\noindent \underline{Note Added:} In a recent paper \cite{FOT}, which
appeared after this paper was completed, the authors consider the
$SO(d,d,Z)$ action on R-R fields from the point of view of low-energy
effective action. One should be able to obtain the same results in our
approach, after determining the $SO(d,d,Z)$ action on gravitinos, and
then using space-time supersymmetry. The results are expected to look
the same as the single T-duality case with $Q_\pm$ and $\O$
appropriately generalized to $SO(d,d)$.
\vspace{-.2cm} 
\section*{Acknowledgments}
I would like to thank C. Angelantonj, I. Antoniadis, A. Dhar, K.
F\"{o}rger, D. Ghoshal, D. Jatkar, B. Pioline, A. Sagnotti, and A. Sen
for many useful discussions during the course of this work.
\vspace{-.2cm} 
\newcommand{\resection}[1]{\setcounter{equation}{0}\section{#1}}
\newcommand{\appsection}[1]{\setcounter{equation}{0}\section*{Appendix}}
\renewcommand{\theequation}{\thesection.\arabic{equation}}
\appendix
\appsection

\resection{${\mathbf \G}$-Matrix Conventions}
We use the metric signature $\{-1, +1,\ldots, +1\}$ and Gamma matrix
conventions  
\beq
\{\G^a,\G^b\}=2\eta^{ab} \,,
\label{G1}
\eeq
so that in the Majorana-Weyl representation all $\G^a$ are real, with 
$\G^0$ antisymmetric and others symmetric. We also need the identity
(with $\epsilon^{01\cdots 9} = 1$)
\beq
\G^{M_1\cdots M_n}\G_{11} = \frac{-(-1)^{n(n-1)/2}}{\sqrt{-G}(10-n)!}
\epsilon^{M_1\cdots M_{10}}\G_{M_{n+1}\cdots M_{10}}\,.
\label{apA-G2}
\eeq
\resection{Type-IIB Supergravity}
For the gravitino and dilatino supersymmetry variations in type-IIB
supergravity, we start with the $SU(1,1)$ invariant formulation of the
theory in \cite{JHS}. Using a prime to indicate the use of the Einstein
metric and after scaling the 5-from field strength to match the
standard string theory conventions, we have
\bea
&&\delta\lambda'= i\G'^MP_M\e'^* - \frac{i}{24}{\G'}^{KLN}G_{KLN}\e'
+\cdots \,,
\\
&&\delta\Psi'_M =D_M\e'+\frac{1}{96}\left(\G'^{\,\,KLN}_M
G_{KLN} -9\G'^{LN}G_{MLN}\right)\e'^* \nonumber\\
&&\qquad\qquad\qquad\qquad\qquad
+\frac{i}{4(480)}\G'^{KLNPQ}\G'_M F_{KLNPQ}\e'+ \cdots\,.
\label{apC-IIBE}
\eea
Here, $\e'$, $\Psi'_M$ and $\lambda'$ are complex Weyl spinors with 
$\G_{11}\e'=\e'$, $\G_{11}\Psi'=\Psi'$, while $\lambda'$ has negative
chirality, and   
\beq
\ba{rclcrcl}
D_M\e' &=&\left(\del_M + \frac{1}{4}w'_{Mab}\G^{ab} 
-\frac{i}{2}Q_M \right)\e'\,,& \qquad &
G_{KLN} &=& -\epsilon_{\al\be}V^\al_+ F^\be_{KLN}\,,
\\
P_M &=& -\epsilon_{\al\be}V^\al_+ \del_M V^\be_+\,,& \qquad &
Q_M &=& -i\epsilon_{\al\be}V^\al_- \del_M V^\be_+\,. 
\ea
\eeq
$\al,\be=1,2$ are $SU(1,1)$ indices, $V^\al_\pm$ is an $SU(1,1)$
matrix $(V^1_-V^2_+-V^1_+V^2_- =1)$ and $F^1_{KLN}=F^{2*}_{KLN}$.
To identify the fields in the usual string theory conventions, we go
to the $SL(2,R)$ formulation by writing $F^\al$ in the real
basis. Then the NS-NS and R-R 2-forms $B_{MN}$ and $C^{(2)}_{MN}$ are
given by   
$$
\left(\ba{c}-dC^{(2)}\\dB \ea\right)=
\left(\ba{c}Re (F^1)\\Im (F^1)\ea\right)=
h \,\left(\ba{c}F^1/\sqrt{2}\\F^2/\sqrt{2}\ea \right)\,,\quad
{\rm with}\,,\quad 
h=\frac{1}{\sqrt{2}}\left(\ba{cc}1&1\\-i&+i\ea \right)\,.
$$
The dilaton and the R-R scalar $C^0$ are identified by parameterizing
the matrix $V$ such that
\beq
U=hV\equiv h \left(\ba{cc}V^1_- & V^1_+ \\ V^2_-& V^2_+ \ea\right)
= \frac{1}{\sqrt{2\tau_2}}\left(\ba{cc} -\bar\tau e^{i\theta}&
 -\tau e^{-i\theta}\\ e^{i\theta} & e^{-i\theta}\ea\right)\,,
\label{apC-UV}
\eeq
with $\tau=C^{(0)}+ie^{-\phi}$. We can set $\theta=0$ by fixing the
$U(1)$. In these conventions, the $SL(2,R)$ action takes the form
\beq
\tau\rightarrow \frac{p\tau+q}{r\tau+s}\,,\qquad 
\left(\ba{c} C^{(2)} \\ B\ea\right)\rightarrow
\left(\ba{cc} p & q \\ r & s \ea\right)
\left(\ba{c} C^{(2)} \\ B\ea\right)\,.
\label{apC-SL2R}
\eeq
$C^{(4)}$ also transforms such that $F^{(5)}$ is invariant (see
(\ref{apC-F})). Having identified the dilaton, we define the string
frame metric and associated spinors as, 
$$
G_{MN}=e^{\phi/2}G'_{MN}\,,\quad
\e = e^{\phi/8}\e'\,,\quad\,
\lambda = e^{-\phi/8}\lambda'\,,\quad
\Psi_M= e^{\phi/8}(\Psi'_M + \frac{i}{4}\G'_M \lambda'^*)\,.
$$
Furthermore, we write the complex Weyl spinors in terms of real
Majorana-Weyl spinors, $\e= \e_+ + i\e_-$, $\Psi_M=\Psi_{+M}+i
\Psi_{-M}$, $\lambda=\lambda_-+i\lambda_+$, where the subscript 
``$\pm$'' is chosen to denote the worldsheet sector that contributes
the spin content of the spinor. The supersymmetry variations
(\ref{apC-IIBE}) then take the form (to linear order in the spinors),
\bea
\hspace{-.9cm}\delta_\pm\lambda_\pm\!\!\!\!&=&\!\!\!\!\frac{1}{2}
\left(\G^M\del_M\phi\mp\frac{1}{12} \G^{M_1M_2M_3}H_{M_1M_2M_3}\right)
\e_\pm + \cdots\,,
\label{apC-dDNS}\\
\hspace{-.9cm}\delta_\mp\lambda_\pm\!\!\!\!&=&\!\!\!\!\frac{1}{2}
e^\phi\left(\pm\G^M F^{(1)}_M + \frac{1}{12}\G^{M_1M_2M_3}
F^{(3)}_{M_1M_2M_3}\right)\e_\mp + \cdots\,,
\label{apC-dDR}\\ 
\hspace{-.9cm}\delta_\pm\Psi_{\pm M}\!\!\!\!&=&\!\!\!\!
\left(\del_M + \frac{1}{4}\,
(w_{Mab}\mp\frac{1}{2}H_{Mab})\,\G^{ab}\right)\e_\pm\,\,+\,
\cdots\,,
\label{apC-dGNS} \\
\hspace{-.9cm}\delta_\mp\Psi_{\pm M}\!\!\!\!&=&\!\!\!\!
\frac{1}{8}e^\phi\Big(\!\!
\mp\G^{M_1}F^{(1)}_{M_1}\!-\!\frac{1}{3!}\G^{M_1M_2M_3}F^{(3)}_{M_1M_2M_3}
\!\mp\!\frac{1}{2(5!)}\G^{M_1\cdots M_5}F^{(5)}_{M_1\cdots M_5}\Big)
\G_M\e_\mp + \cdots\,.
\label{apC-dGR}
\eea
The R-R potentials $C^{(n)}$ are defined such that,
\beq
F^{(n)}_{M_1\cdots M_n} = n\del_{[M_1} C^{(n-1)}_{M_2\cdots M_n]}
-\frac{n!}{3!(n-3)!}H_{[M_1M_2M_3} C^{(n-3)}_{M_4\cdots M_n]}\,.
\label{apC-F}
\eeq

\resection{Type-IIA Supergravity}
The gravitino and dilatino supersymmetry variations in type-IIA theory
are given in \cite{LR} for massive IIA. When written in terms of
appropriate variables, they lead to the usual massless IIA equations
when the mass parameter is set to zero. In the standard string theory
normalizations for the fields, these equations take the form,
\bea
\delta\lambda'\!\!\!\!&=&\!\!\!\!\frac{1}{2}\Big[\G'^M\del_M\phi-\frac{1}{12}
\G'^{MNP}H_{MNP}\G_{11}\Big]\e'\nonumber\\
&&\!\!\!\!+\frac{1}{8}\Big[5 e^{5\phi/4} F^{(0)}-\frac{3}{2!}e^{3\phi/4}
\G'^{MN}F^{(2)}_{MN}\G_{11}+\frac{1}{4!}e^{\phi/4}\G'^{MNPQ}
F^{(4)}_{MNPQ}\Big]\e'+\cdots\,,
\\
\delta\Psi'_M\!\!\!\!&=&\!\!\!\!\Big[\del_M +\frac{1}{4}w'_{Mab}\G^{ab}
+ \frac{1}{96}e^{-\phi/2}\left(\G_M'^{\,\,NPQ}-9\delta^{[N}_M\G'^{PQ]}
\right)H_{NPQ}\G_{11}\Big]\e'
\nonumber \\
&&\!\!\!\!+\frac{1}{32}\Big[-\frac{1}{2}e^{5\phi/4}\G'_M F^{(0)}-
\frac{1}{2}e^{3\phi/4}\left(\G_M'^{\,\,NP}-14\delta^{[N}_M\G'^{P]}
\right)F^{(2)}_{NP}\G_{11}\nonumber\\ 
&&\!\!\!\!
+\frac{1}{4}e^{\phi/4}\left(\G_M'^{\,\,NPQR}-\frac{20}{3}\delta^{[N}_M
\G'^{PQR]}\right)F^{(4)}_{NPQR}\Big]\e'+\cdots\,. 
\eea
Here, a prime indicates the use of the Einstein metric, ``$\cdots$''
denote 3-spinor terms and the field strengths $F^{(n)}$ are given by  
\beq
\ba{l}
F^{(0)}=m\,,\\
F^{(2)}_{MN}=2\del_{[M}C^{(1)}_{N]} + m B_{MN}\,,\\
F^{(4)}_{MNPQ}= 4\del_{[M}C^{(3)}_{NPQ]} - 4H_{[MNP}C^{(1)}_{Q]}
+ 3m B_{[MN} B_{PQ]}\,.
\ea
\label{apD-Fm}
\eeq
The constant $m$ is the mass parameter of the massive type-IIA theory
and the usual massless IIA theory is recovered by setting $m=0$, in
which case the these equations take the form (\ref{apC-F}) above. The
string frame metric and spinors are given by,
$$
G_{MN}=e^{\phi/2}G'_{MN}\,,\quad
\e = e^{\phi/8}\e'\,,\quad\,
\lambda = e^{-\phi/8}\lambda'\,,\quad
\Psi_M= e^{\phi/8}(\Psi'_M + \frac{1}{4}\G'_M \lambda')\,.
$$
Let us consider the above equations in terms of the positive and
negative chirality components of $\e$ and other spinors. One can then
see that the type-IIA theory described in \cite{LR} is the one in
which the positive chirality component of $\e$ originates in the
left-moving worldsheet sector. However, in our conventions for
T-duality, we need the IIA in which the positive chirality component
of $\e$ originates in the right-moving worldsheet sector. This IIA
theory is obtained from the one described in \cite{LR} by a
worldsheet parity transformation that reverses the signs of $H_{MNP}$ 
and $F^{(2)}$, keeping $F^{(0)}$ and $F^{(4)}$ unchanged. Then the
above equations lead to,
\bea
\hspace{-.9cm}\delta_\mp\lambda_\pm\!\!\!\!&=&\!\!\!\!\frac{1}{8}
e^\phi\left(5 F^{(0)} \pm \frac{3}{2!}\G^{M_1M_2} F^{(2)}_{M_1M_2} 
+ \frac{1}{4!}\G^{M_1M_2M_3M_4}F^{(4)}_{M_1M_2M_3M_4}\right)\e_\mp
+\cdots\,,
\label{apD-dDR}\\ 
\hspace{-.9cm}
\delta_\mp\Psi_{\pm M}\!\!\!\!&=&\!\!\!\!\frac{1}{8}e^\phi
\Big[F^{(0)}\pm\frac{1}{2!}\G^{M_1M_2}F^{(2)}_{M_1M_2}+
\frac{1}{4!}\G^{M_1M_2M_3M_4}F^{(4)}_{M_1M_2M_3M_4}\Big]
\G_M\e_\mp +\cdots\,.
\label{apD-dGR}
\eea
The variations $\delta_\pm\lambda_\pm$ and $\delta_\pm\Psi_{\pm M}$
are still given by equations (\ref{apC-dDNS}) and (\ref{apC-dGNS})
though now, $\e_-$, $\Psi_{-M}$ and $\lambda_+$ have positive
chirality and $\e_+$, $\Psi_{+M}$ and $\lambda_-$ have negative
space-time chirality.

\noindent\underline{Supersymmetry Variations of NS-NS Fields:}
\beq
\delta_\pm G_{MN}= 2\,\bar\e_\pm\,\G_{(M}\Psi_{\pm N)}\,,\qquad
\delta_\pm B_{MN}=\pm 2\,\bar\Psi_{\pm[M}\,\G_{N]}\e_\pm\,,\qquad
\delta_\pm \phi= \bar\lambda_\pm \e_\pm\,.
\label{ap-dNS}
\eeq 
Here, $G_{MN}$ is the string metric and ${}^{_(}{}^{_)}$ denotes
symmetrization with unit weight. These equations are valid in both IIA
and IIB.

\end{document}